\documentclass{acm_proc_article-sp}

\usepackage{paralist}
\usepackage{graphicx}
\usepackage{xcolor,colortbl}

\begin{document}

\title{Where Do We Stand in Requirements Engineering Improvement Today? First Results from a Mapping Study}

\numberofauthors{4} %  
\author{
% 1st. author
\alignauthor
Daniel M\'{e}ndez Fern\'{a}ndez, Saahil Ognawala\\
       \affaddr{TU M\"unchen, Garching}\\
       \email{mendezfe,ognawala@in.tum.de}
% 3rd. author
\alignauthor Stefan Wagner\\
       \affaddr{University of Stuttgart}\\
       \email{stefan.wagner@informatik.uni-stuttgart.de}
\and
% 4th. author
\alignauthor Maya Daneva\\
       \affaddr{University of Twente}\\
       \email{m.daneva@utwente.nl}
}

\maketitle
\begin{abstract}
Requirements engineering process improvement (REPI) approaches have gained much attention in research and practice. 
% Problem Statement
So far, however, there is no comprehensive view on the research in REPI in terms of solutions and current state of reported evidence.
% Research Method
This paper aims to provide an overview on the existing solutions, their underlying principles and their research type facets, i.e.\ their state of empirical evidence. To this end, we conducted a systematic mapping study of the REPI publication space. 
% Results
This paper reports on the first findings regarding research type facets of the contributions as well as selected
methodological principles. We found a strong focus in the existing research on solution
proposals for REPI approaches that concentrate on normative assessments and benchmarks of the RE activities rather than on holistic RE improvements according to individual goals of companies.
% Conclusions
We conclude, so far, that there is a need to broaden the work and to investigate more problem-driven REPI which also targets the improvement of the quality of the underlying RE artefacts.
\end{abstract}

\keywords{Requirements Engineering Process Improvement, Systematic Mapping Study}

\section{Introduction}
\label{Introduction}

Requirements engineering (RE) aims at the discovery and specification of requirements that unambiguously reflect the purpose of a software system. Thus, RE is an important factor for productivity and quality. Given the practical importance of RE, it remains a complex discipline driven by uncertainty~\cite{MWLBC10} which eventually makes RE hard to investigate and even harder to improve~\cite{MW13}. 
Even though a significant number of contributions have been made in the research field of requirements engineering process improvement (REPI), we do not have exhaustive knowledge about the proposed solutions, the problems they address and the state of evaluation and validation of these solutions. There exist secondary studies that deal with the larger context of SPI but none so far for improving RE concerning all its particularities.  We aim to consolidate the current understanding about the state-of-the-art by conducting a systematic mapping study of all publications on RE process improvement. In this paper, we report on our results and focus, as a first step, on categories of publications according to research type facets, the contribution phases, paradigms and their underlying principles. The complete data and analysis can be found in~\cite{OMW14}.

\section{Study Design}
\label{StudyDesign}
Our study design follows the standard procedures of a systematic mapping study \cite{PFMM08}. We did this in conjunction with the methods of a systematic literature review which entails a further in-depth analysis of selected publications.

\subsection{Research Questions}
To systematically describe the state-of-the-art, we will answer the following research questions on REPI publications. 

\textbf{RQ1: Of what type is the research?} As a first step, we will classify the REPI publications according to the research type facets as described by Wieringa et al.\ \cite{WMMR06}. A research type facet is an abstract description of the activity stage in the engineering cycle that is in scope of a contribution. We also aim to spot trends in the facets of REPI papers over the years. We list the available research type facet categories in Tab.~\ref{tab:rq4choices}. 

\begin{table}[h]
\centering
\caption{Definitions of research type facets~\cite{WMMR06}.}
\footnotesize
\begin{tabular}{p{43pt}p{170pt}}
\hline
\textbf{Validation paper}          & Techniques investigated are novel and have not yet been implemented 
			  in a large scale industrial or academic setting. \\ \hline
\textbf{Evaluation paper}          & Techniques are implemented and evaluated in a large scale industrial, academic or other real world setting.                                                                                                                 \\ \hline
\textbf{Solution proposal}   & A solution to a problem is proposed, either novel or an extension to an existing solution. \\ \hline
\textbf{Philoso\-phical paper} & It proposes a new way of looking at existing problems by re-structuring the field in form of a taxonomy, conceptual framework or systematic literature review.                                                                                                                                    \\ \hline
\textbf{Opinion paper}       & The authors present their opinion on a problem space with a critical view on one or more solutions described by other researchers trying to tackle the problem. \\ \hline
\textbf{Experience paper}    & It provides a retrospective view on the authors' experience in developing, applying and evaluating a certain technique in the field of engineering process improvement.  \\ \hline
\textbf{Explo\-ra\-tory paper}   & It deals mainly with the problem space with a bird's eye view of the common problems faced by various solutions proposed.
\\
\hline
\end{tabular}
\label{tab:rq4choices}
\end{table}

\textbf{RQ2: Which process improvement phases are considered?} Having classified the overall contributions according to their facet, we want to know whether those contributions take a holistic view on REPI or whether they focus on selected improvement phases only. We distinguish between (a) \emph{Analysis} where the focus lies on analysis and assessment of a RE, (b) \emph{Construction}  where the focus lies lies on the (re-)design of a RE process and, thus, on the actual improvement realisation, (c) \emph{Validation} where the focus lies on the validation of an improvement endeavour, and (d) \emph{RE Process Improvement Lifecycle (REPI-LC)} where the contribution takes a holistic view on all phases and/or on general metrics and measurements.

\textbf{RQ 3: What paradigms do the publications focus on?} 
We distinguish between activity-oriented and artefact-orien\-ted paradigms based on whether the publications focus on improving the quality of the activities that form a part of the RE processes or on improving the quality of the RE artefacts created. If contributions do not differentiate between the paradigms (e.g.\ when providing a set of metrics and measurements) or simply include ideas dealing with none of the two paradigms in particular, then we will not assign the contribution any paradigm focus. 

\textbf{RQ 4: Are the underlying principles of normative or of problem-driven nature?} 
We categorise a publication either as normative, where a given RE activity/artefact is assessed and improved against a given external norm, or as problem-driven where the improvement is conducted against company-specific goals and problems (see also~\cite{PIGO08}). 

\subsection{Study Selection}
As Petersen et al.~\cite{PFMM08} recommended, we started our mapping study with an exhaustive search of a publication database with the key concept terms in REPI. We did a pre-analysis of a selected set of key publications in the REPI area and made a map of the chief search terms that seemed closely related to these papers. Next, we performed snowballing on the selected publications as suggested by Kitchenham et al.~\cite{KPBTBL09}. This gave us a large initial dataset with a list of key publications and the main concept keywords. We can now form the search query strings and modify them based on the quality of the search result set (as compared to the initial dataset). 

\subsection{Data Collection Procedures}
Our data collection procedure is an automated search on established web databases including \emph{ACM Digital Library, SpringerLink, ScienceDirect} and \emph{IEEE Xplore}.

We use the keywords present in the initial dataset to define search query terms. Lists of prominent contributors in the domain and their publications are a control mechanism to filter out irrelevant search results and tweak the search string correspondingly.  Another set of notable additions to the contribution data are technical reports and academic studies in form of PhD theses which often do not form a part of the search result space in the above listed sources. We search for such contributions using Google Scholar which has a wider span that indexes titles located in repositories like university databases and other independent publications. 

\subsection{Inclusion and Exclusion Criteria}
Once we have  a set of contributions from the publication databases, we use a list of inclusion (IC) and exclusion (EC) criteria (described in Tab.~\ref{tab:inclusionexclusion}) on this dataset before the analysis and voting stage. 
\vspace{-2mm}

\begin{table}[htb]
\centering
\caption{Inclusion and exclusion criteria.}
\footnotesize
\begin{tabular}{p{15pt}p{190pt}}
\hline
$IC_{1}$ & The paper directly relates to REPI.\\
$IC_{2}$ & The title and abstract refer to REPI.\\
$IC_{3}$ & The keywords contain related words.\\
$IC_{4}$ & The contribution addresses the research questions, i.e.\ it\\
    & \ldots introduces, discusses, compares, or evaluates\ldots\\   
    & \ldots approaches or experiences, terms and concepts and/or metrics to\ldots \\
    & \ldots improve (assess and/or implement and/or evaluate) requirements engineering processes or artefacts.\\
\hline
$EC_{1}$ & The paper addresses SPI in general (without clear linkage to RE).\\
$EC_{2}$ & The topic does not address approaches, studies or experiences for improving requirements
engineering but new approaches and techniques that are claimed to improve RE as an effect
of applying them (e.g.\ elicitation techniques).\\
$EC_{3}$ & No scientific publication, i.e.\ PowerPoint presentations, abstracts or posters.\\
$EC_{4}$ & The contribution's language is not English.\\
$EC_{5}$ & The contribution is not available.\\
$EC_{6}$ & The contribution appears multiple times in the result set (see below).\\
$EC_{7}$ & The contribution investigates (industrial) problems in RE to be addressed by 
research to improve RE.                                                                                       
\\ [1pt] \hline
\end{tabular}
\label{tab:inclusionexclusion}
\end{table}

Among contributions where the same approach is reported, we only choose one to include in our study; e.g.\ PhD theses forming a cumulative report of various approaches. 
We treat papers where several techniques or approaches are reported as a single contribution. 
Systematic literature reviews are treated as \emph{philosophical papers}~\cite{WMMR06} because they define and organise existing concepts and approaches taking a novel view. 
In case of metrics being introduced in a paper that can be applied to both artefact or activity orientation, we set the paradigm to "N/A". 
Table~\ref{tab:dataset} gives numbers of papers at each stage of data processing. \emph{Results seen} are all the results returned by the database search in the top 20 pages. \emph{Included} papers are the ones that were kept after filtering the seen results in the analysis stage by making use of the inclusion and exclusion criteria in  Tab.~\ref{tab:inclusionexclusion}. We then undergo the first round of voting where we further filter out more publications based on their relevance to our research questions so as to derive the \emph{Relevant} result set. \\ 
\vspace{-3mm}
\begin{table}[htb]
\centering
\caption{Dataset summary.}
\footnotesize
\begin{tabular}{p{50pt} r r r r}
\hline
Database  & Total & Results &  &  \\
Name  & results & seen & Included & Relevant \\ [1pt] \hline
ACM            & 81            & 81           & 23       & 15       \\
SpringerLink   & 349           & 349          & 31       & 11       \\
ScienceDirect  & 132           & 132          & 12       & 2        \\
Google Scholar & 276           & 276          & 16       & 11       \\
IEEE Explore   & 2,819,217       & 275          & 18       & 15       \\
Misc.          & 4             & 4            & 4        & 4        \\ [1pt] \hline
Total          & 2,820,059       & 1117         & 104      & 58       \\ [1pt] \hline
\end{tabular}
\label{tab:dataset}
\end{table}

\subsection{Analysis and Voting Procedure}
We did a staged voting procedure on the set of 58 papers, as indicated in Tab.~\ref{tab:dataset}. Each senior researcher (Daneva, M\'{e}ndez and Wagner) worked individually to categorise every publication according to the research type facets, the lifecycle phase, the two paradigms and the two underlying principles. The voting procedure allowed us to put forward arguments regarding our respective choices on assigning a paper to a category of relevance to the RQs. Once each researcher's individual categorisation was over, the researchers got together to compare and contrast their categorisations until reaching agreement (occasionally based on in-depth analysis of the paper). Each voting stage ended with a consensus-driven discussion among the three researchers on assigning categories to a paper on which there were disagreements.

We conducted the voting procedure over four stages yielding at each stage the agreement level in the classification subsequently shown:

\begin{compactenum}
\item[$1^{st}$ stage:] 53.4~\% (31/58)
\item[$2^{st}$ stage:] 72.4~\% (42/58) 
\item[$3^{st}$ stage:] 86.2~\% (50/58)  
\item[$4^{st}$ stage:] 100~\%  (58/58)  
\end{compactenum}

\begin{figure*}[hbt]
\centering
\includegraphics[width=1\textwidth]{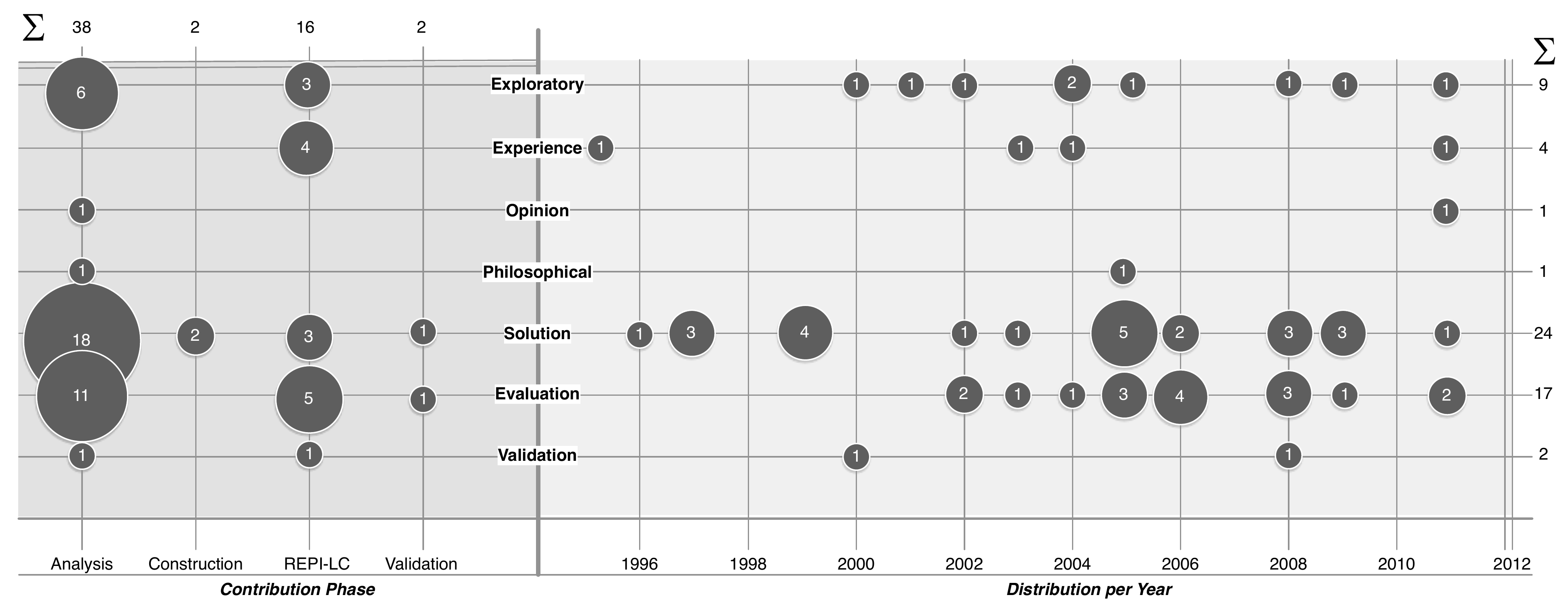}
\caption{Facet distribution according to years and lifecycle phases.}
\label{fig:facetphaseyear}
\end{figure*}

\section{Results}
\label{Results}
We present our findings structured according to our RQs.

\subsection{Research Type Facet (RQ 1)}
Fig.\ \ref{fig:facetphaseyear}, right side, shows the distribution of the contribution over the years and according to their research type facets. 41 out of 58 publications analysed were \emph{solution} proposals or \emph{evaluation} papers. We can see a lack of retrospective analysis in the form of experience reports which we consider important given that especially RE and, thus, its improvement strongly depends on subjectivity (beliefs, desires, fears, experiences and expectations), but in general a regular distribution over solution proposals and evaluation papers. Overall, the map suggests the beginning of research in this area on basis of Sawyer et al.~\cite{SSV97} introducing the REPI based on a set of what they consider to be best practices. The results also indicate that most of the proposed REPI approaches focus on extending the basic ideas introduced by Sawyer et al. and grounding REPI on best practices (see also subsequent RQs).

\subsection{Phase of contribution (RQ 2)}
38 out of the 58 publications fall under the \emph{analysis phase} of the REPI lifecycle. Fig.\ \ref{fig:facetphaseyear}, left side, shows the distribution of the publications over all the phases and depicts which research type facet are in scope. The map suggests that most of the papers focus on the analysis phase while little seems yet proposed for the contraction phase, i.e. the realisation of actual improvement forecasts identified in an analysis phase.

\subsection{Contribution Paradigm (RQ 3)}
We found that 48 out of 58 papers presented the activity-oriented paradigm as the one adopted in REPI. In 7 out of the 58 papers we could not find enough evidence indicating the kind of paradigm adopted as the focus was, for example, on metrics and measurements used in various REPI phases. However, it remained unclear if measurements supported activity oriented or artefact-oriented REPI. The focus of most contributions lies on normative improvements focusing on the RE activities carried out. This is in tune with our observation that many contributions are an extension of the initial work proposed by Sawyer et al., thus, those contributions focus on how to assess RE processes against a given norm consisting of proposed RE best practices considered to comprise an external notion of ``good RE''.

\subsection{Contribution Principle (RQ 4)}
\begin{figure}[htb]
\centering
\includegraphics[width=.6\columnwidth]{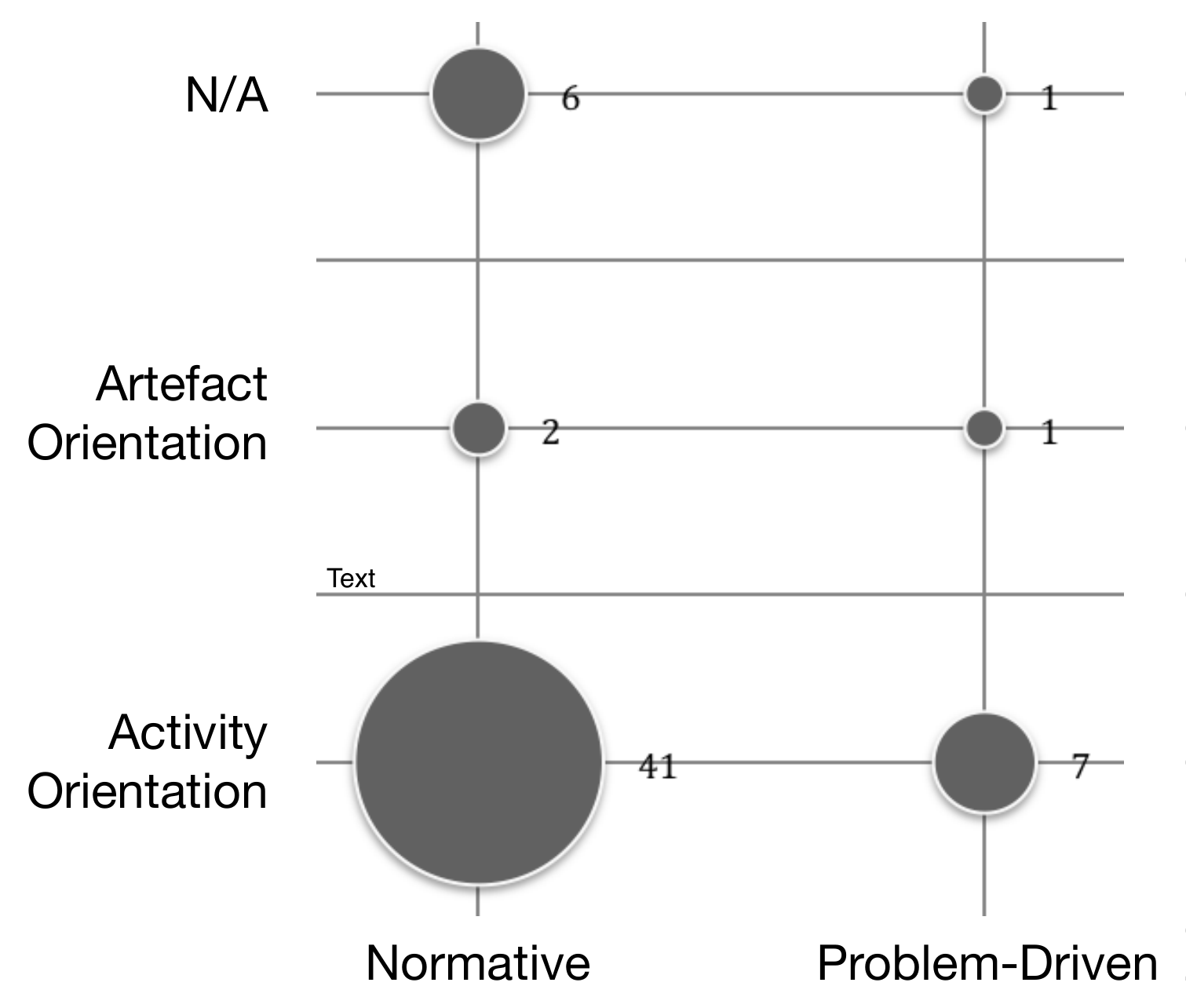}
\caption{Paradigms and principles.}
\label{fig:paradigm-principle}
\end{figure}

Fig.~\ref{fig:paradigm-principle} reports our findings on the classification of the principle and maps them against the paradigms (RQ 2). 
41 out of 58 papers fall in the \emph{normative} category where the focus lies mostly on improving activities rather than RE artefacts. The papers in this category mostly deal with sets of best practices as external norms where the current state of RE in a company is assessed and aligned with those best practices. 

\section{Conclusion}
\label{Conclusion}
In this paper, we presented first results from a mapping study on the current state of REPI. Based on 58 primary studies, we showed that most contributions focus on the improvement of RE activities while the improvement of RE artefacts is barely discussed. In addition, most improvement approaches focus on a normative improvement where external norms of best practices are taken as a reference. Considering that most contributions focus on assessments rather than on realising an improvement, we conclude that available contributions provide means to rate and assess the current state of RE in companies against an external norms of activity-centric best practices of which most arise from initial contributions made by Sawyer et al.~\cite{SSV97}. 

Our study further revealed that very few exploratory papers have been published in REPI. This means we have relatively little evidence (and hence, knowledge) about the full range of problems that organisations face. In turn, the general applicability of the solution proposals might well be compromised. For example, as currently RE is applied more and more to new domains (e.g. smart city systems), we think it would be unrealistic to assume that the solutions proposals would catch up with the REPI needs of organisations executing projects in those domains. We therefore call for more exploratory studies in REPI to identify and better understand common problems faced.  Next, we found very few (3 out of 58) papers taking an artefact-oriented perspective on REPI. Clearly, aspects such as effectiveness and efficiency of RE activities are related to activities and this might explain the massive amount of papers adopting activity-oriented paradigm for REPI. However, are the RE activities actually problematic in the real world? In fact, empirical RE papers report of problematic artefacts (e.g.\ specifications, models). We assume  that the activity-oriented paradigm is studied so often because it has established itself through the ``best-practice'' movement. Last, we found the majority of papers were of normative nature. It is tempting to assume this finding is traceable to the established ``best-practice'' thinking in the software industry. However, even best practice gurus (e.g.\ Capers Jones) suggest that problem-driven improvement might yield greater benefits than a big-bang best-practice based approach. 

We therefore think that investing in problem-driven REPI that also considers the quality of RE artefacts would be worthwhile and necessary to fully understand the broad spectrum of REPI possibilities.

\paragraph{Limitations}

There are two main limitations of this mapping study: First, the possible bias in the selection of papers for inclusion as our access to relevant sources depended on the  appropriateness of the used search strings. In the REPI area, a broad diversity of  terms is used which implies a risk that we might have missed some relevant studies. We took extra steps to counter this risk by analysing keywords and publications of leading REPI authors. 

Second, it might be possible that we collectively categorised a paper in a wrong way. We countered this by implementing a four-stage voting procedure focused on argumentation, repeated reviews and consensus building. We therefore think the risk of this threat is minimal. Yet, the RE paper classification in \cite{WMMR06} has not been created with systematic reviews in mind, and at times we found it difficult to categorise a paper to only one facet. 

\bibliographystyle{abbrv}
\bibliography{esem}  

\newpage
Copyright ACM. This is the author's version of the work. It is posted here for your personal use. Not for redistribution. The definitive Version of Record was published in Proceedings of the 8th ACM/IEEE International Symposium on Empirical Software Engineering and Measurement, http://dx.doi.org/10.1145/2652524.2652555.

\end{document}